\documentclass[twocolumn]{aa}
\usepackage{graphicx}
\usepackage{times}
\usepackage{natbib}
\usepackage{epsfig}
\bibpunct{(}{)}{;}{a}{}{,}

\usepackage{color}

\usepackage{amssymb}
\usepackage{color}
\begin{document}

\title{Full characterization of binary-lens event OGLE-2002-BLG-069 from PLANET observations 
       \thanks{Based on observations made at ESO, 69.D-0261(A), 269.D-5042(A), 169.C-0510(A)  }}
  \titlerunning{OGLE-2002-BLG-069 model}
  \authorrunning{D.~Kubas et al. }

\author{ 
    D.~Kubas\inst{1,7},        A.~Cassan\inst{1,2},  J.P.~Beaulieu\inst{1,2},
    C.~Coutures\inst{1,2,4}, M.~Dominik\inst{1,5}, M.D.~Albrow\inst{1,10}, 
    S.~Brillant\inst{1,3},  J.A.R.~Caldwell\inst{1,11}, D.~Dominis\inst{1,7},            
    J.~Donatowicz\inst{1,6}, C.~Fendt\inst{7},    P.~Fouqu\'e\inst{1,12},     
    U.G.~J{\o}rgensen\inst{1,9},     
               J.~Greenhill\inst{1,13},
    K.~Hill\inst{1,13},     J.~Heinm\"uller\inst{7},   K.~Horne\inst{1,5},
    S.~Kane\inst{1,5}, J. B.~Marquette\inst{2}, R.~Martin\inst{1,13},
    J.~Menzies\inst{1,14},            K.R.~Pollard\inst{1,10},
    K.C.~Sahu\inst{1,11},             C.~Vinter\inst{1,9}, 
    J.~Wambsganss\inst{1,8},          R.~Watson\inst{1,13},
    A.~Williams\inst{1,15}
    \and C.~Thurl\inst{16}
  }

  \offprints{kubas@astro.physik.uni-potsdam.de}
   \institute{
{PLANET collaboration member} 
\and
{Institut d'Astrophysique de Paris, 98bis Boulevard Arago, 75014 Paris,
France} 
\and
{European Southern Observatory, Casilla 19001, Vitacura 19, Santiago,
Chile} 
\and
{DSM/DAPNIA, CEA Saclay, 91191 Gif-sur-Yvette cedex, France} 
\and
{University of St Andrews, School of Physics \& Astronomy, North Haugh, St
Andrews, KY16~9SS, United Kingdom} 
\and
{Technical University of Vienna, Dept. of Computing, Wiedner Hauptstrasse
10, Vienna, Austria} 
\and
{Universit\"at Potsdam, Astrophysik, Am Neuen Palais 10, 14469 Potsdam,
Germany} 
\and
{Astronomisches Rechen-Institut, M\"onchhofstrasse 12-14, 69120 Heidelberg, Germany}
\and
{Niels Bohr Institute, Astronomical Observatory, Juliane Maries Vej 30,
 2100 Copenhagen, Denmark} 
\and
{University of Canterbury, Department of Physics \& Astronomy, Private Bag
4800, Christchurch, New Zealand } 
\and
{Space Telescope Science Institute, 3700 San Martin Drive, Baltimore, MD
21218, USA} 
\and
{Observatoire Midi-Pyr\' en\' ees, 14 avenue Edouard Belin, F-31400 Toulouse,
France} 
\and
{University of Tasmania, Physics Department, GPO 252C, Hobart, Tasmania
7001, Australia} 
\and
{South African Astronomical Observatory, P.O. Box 9, Observatory 7935,
South Africa} 
\and
{Perth Observatory, Walnut Road, Bickley, Perth 6076, Australia} 
\and
{RSAA, Mount Stromlo and Siding Spring Observatories, ANU, Cotter Road,
Weston Creek, Canberra, ACT 2611, Australia} 
}
 
  \date{Received ; accepted}
   
   \abstract{We analyze the photometric data obtained by PLANET and OGLE on the caustic-crossing binary-lens
            microlensing event OGLE-2002-BLG-069. Thanks to the  excellent photometric and
            spectroscopic coverage of the event, we are able to constrain the lens model up to the 
            known ambiguity between close and wide binary lenses. The detection of annual parallax in
            combination with measurements of extended-source effects allows us to determine the
            mass, distance and velocity of the lens components for the competing models. While the model
            involving a close binary lens leads to a Bulge-Disc lens scenario with a lens mass of  
            $M=(0.51 \pm 0.15) ~M_\odot$ and distance of $D_{\rm{L}}=(2.9\pm 0.4) ~{\rm{kpc}}$, 
            the wide binary lens solution requires a rather 
            implausible binary black-hole lens ($M \gtrsim 126 ~M_\odot$). 
            Furthermore we compare current state-of-the-art numerical and empirical models for 
            the surface brightness profile of the source, a G5III Bulge giant. We find that 
            a linear limb-darkening model for the atmosphere of the source star is consistent with
            the data whereas a PHOENIX atmosphere model assuming LTE and with no free parameter
            does not match our observations.

}

  \maketitle
%
\section{Introduction} \label{sec:intro}
%

By exploiting the phenomenon of the bending of light from background source stars due 
to the gravitational field of intervening compact objects acting as lenses, Galactic 
microlensing provides an opportunity to infer the brightness profile
of the source star, the mass and configuration of the lens, as well as the 
relative parallax and proper motion.
In recent years there has been a remarkable increase in the power of microlensing 
survey alert systems like OGLE-III \citep{Udal03}\footnote{www.astrouw.edu.pl/$\sim$ogle/}
and MOA \citep{2001MNRAS.327..868B}\footnote{www.physics.auckland.ac.nz/moa/}. 
As a consequence, binary-lens microlensing events have become a unique and valuable
tool to study, in unprecedented detail, members of the source and lens population within our Galaxy and in 
the Magellanic Clouds
\citep{Abe2003,Fields2003,An2002,Albrow2000c,Albrow2000b,Albrow2000a,Albrow1999,Albrow1999b,Albrow1999a}.
The OGLE-2002-BLG-069 event is an ideal example for showing the current capabilities of microlensing 
follow-up observations.\\
The passage of a source star over a line-shaped (fold) caustic as created by a binary 
lens produces a characteristic peak in the light curve which depends on 
the stellar brightness profile. 
The data obtained for OGLE-2002-BLG-069 clearly reveal a pair of such passages 
consisting of an entry and subsequent caustic exit, where the number of
images increases by two while the source is inside the caustic.\\  
This binary-lens event is the first where both photometric and high-resolution 
spectroscopic data were taken over the whole course of the caustic exit.
 The previous attempts on EROS-2000-BLG-5 \citep{2001A&A...378.1014A}
had good coverage but low spectral resolution 
\citep{Albrow2001}, or a pair of spectra taken with high 
resolution but low signal-to-noise \citep{Castro2001}. 
Prior to this study, we presented a fold-caustic model of the
OGLE-2002-BLG-069 photometric data comparing a linear law  and a  model derived from
PHOENIX v2.6 synthetic spectra for the limb-darkening and
analyzed variations in the $\element{H}\alpha$ line as observed in high-resolution
UVES spectra taken over the course of the caustic passage \citep{2004A&A...419L...1C}. 
A full account of the spectral observations 
in  $\element{H} \alpha,~\element{H} \beta$, $\element{CaII}$, $\element{Mg}$ 
and other lines will be given in \cite{Beaulieu2004}. Here, we concentrate on 
the photometric data alone 
in order to present the full binary-lens model.\\
For the majority of observed microlens events all information about lens mass, relative 
lens-source distance and proper motion is convolved into one single characteristic time scale. 
Binary-lens events however are especially sensitive to effects caused by finite source size and parallax, so that
in combination with the determination of the angular source radius, these three lens quantities can be measured 
individually \citep{1966MNRAS.134..315R,1992ApJ...392..442G}.    
This is only the second binary  microlensing event, after EROS-BLG-2000-5 \citep{2002ApJ...572..521A},
for which this has been achieved. 
Despite our high sampling rate and the small uncertainty of our
photometric measurements, we still encounter the well known
close/wide-binary ambiguity originating in the lens equation itself \citep{1999A&A...349..108D} 
and which may only be broken with additional astrometric measurements 
as proposed in \cite{2001glrp.conf..259D} and \cite{2000ApJ...538..653G}.

%
\section{OGLE-2002-BLG-069 photometry data} \label{sec:photo}
%

\begin{figure}
\includegraphics[width=9.4cm]{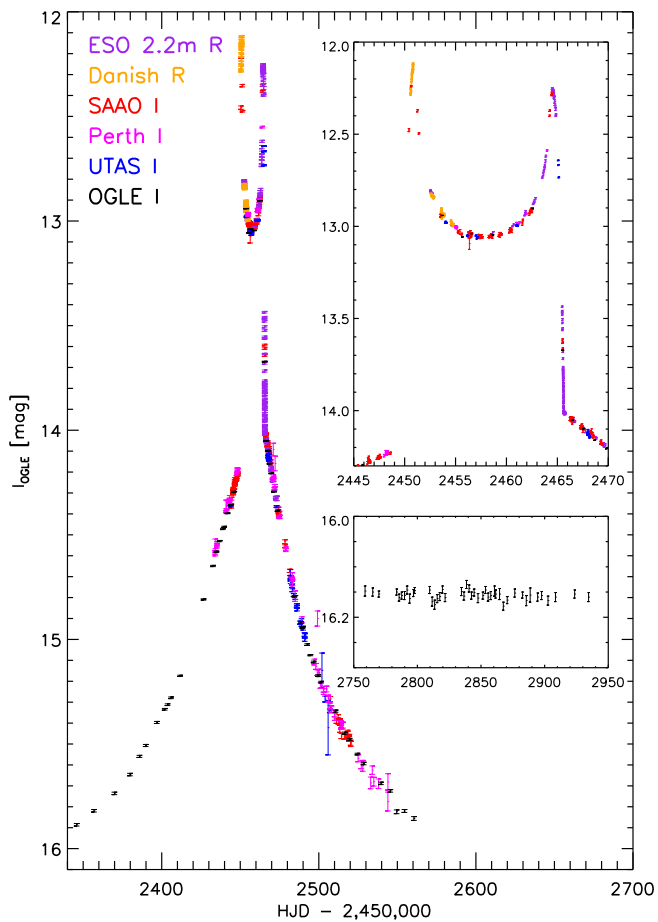}
\caption{Complete photometric I- and R-band datasets of PLANET and OGLE, with PLANET 
calibrated to the OGLE filter. The {\bf{upper inner panel}} shows a zoom of the caustic passages,
while the {\bf{lower panel}} displays the baseline measurements of OGLE made in 
the 2003 season, i.e. one year after the lensing event occurred. }
\label{fig:alldata}
\end{figure}

Alerted by the OGLE collaboration \citep{Udal03} on June 1 2002 about the ongoing 
Bulge microlensing OGLE~2002-BLG-069 event (R.A.=$17^{h}48^{m}1^{s}.0$
, decl.= $-21^{\circ}16^{\prime}9^{\prime\prime}.3$),
the PLANET collaboration network began photometric observations on June 18,
using 6~different telescopes, namely SAAO 1m (South Africa), Danish 1.54m (La Silla),
ESO 2.2m (La Silla), Canopus 1m (Tasmania), Stromlo 50'' (Australia) and Perth 1m (Australia).
Data were taking in $I$- (UTas, Danish, SAAO, Perth), 
$R$- (La Silla)  and $V$-bands
(Stromlo). Since the $V$-band data set of Stromlo contains only 8 points, 
which is less than the number of parameters we fit, we do not use it in the modeling process.\\
The photometry reductions were done by 
point-spread-function (PSF) fitting using our own modified version of DoPHOT \citep{1993PASP..105.1342S},
implemented as part of the PLANET reduction pipeline.
The full raw data set including the public OGLE data (available from www.astrouw.edu.pl/$\sim$ogle/ogle3/ews/ews.html)
consists of 675 points. Data that were obviously wrong according to the observational log books or for which
the reduction software did not succeed in producing a proper photometric
measurement have been eliminated. Moreover, PLANET data taken under
reported seeing that was significantly above the typical value for the
given site were removed according to the cut-offs listed in Table \ref{tab.cut}. Altogether
about 2\,\% of the data were rejected, leaving us with a total of 651 points (Fig.\ref{fig:alldata}).\\
\begin{table}[h!]
\begin{center}
\begin{tabular}{|l|l|l|l|}  \hline
telescope & median seeing & seeing cut & number of \\
 &(arcsec)& (arcsec)& points\\\hline
ESO 2.2m & 1.13& $\leq$ 2.5 & 150  \\   
Danish  & 1.62&$\leq$ 2.5 & 108 \\
UTas &3.13&$\leq$ 3.6 & 58\\
SAAO &1.93 & $\leq$ 2.6 & 153\\
Perth &2.43 &$\leq$ 2.8 & 86\\\hline
\end{tabular}
\caption{Selection criteria for PLANET photometric dataset. }
\label{tab.cut}
\end{center}
\end{table}
%
Thanks to the favourable brightness at baseline ($I_{\rm{OGLE}}= 16.15 \pm 0.01 $ mag)
and low crowding of this event, 
the correlation between seeing and observed flux is negligible and does not yield 
a significant signature in the data, contrary to some previously analyzed events 
\citep[e.g.][]{Albrow2000a}.

\section{Binary-lens model} \label{sec:method}
%

\subsection{Parametrization and general approach}\label{sec:psbl}

The art of binary-lens light curve modeling still poses significant challenges.
A priori the parameters have a large dynamical range ( the mass ratio for instance
can vary over several magnitudes) and the standard $\chi^2$ goodness-of-fit measure of the   
complicated high-dimensional parameter space is very sensitive to subtle changes 
in most of the parameters. Furthermore the parameter space contains both  real ambiguous solutions and false numerical minima 
where parts of the $\chi^2$ surface are  either flat or contain very irregular and rough regions, 
in which local optimizing codes, based on gradient search algorithms \citep{1992nrfa.book.....P}, can get trapped.
Another barrier is the treatment of extended-source effects, which are prominent when the source 
is resolved by caustic crossings.
Given the large number of initial model parameters and the large number of data points,
all suggested methods of calculate binary-lens curves of extended 
sources \citep{1988A&A...191...39K,1995A&AS..109..597D,
1997MNRAS.284..172W,1998A&A...333L..79D}
are too time consuming to be used for the complete data set.
Therefore we employ a strategy similar to that used by \cite{Albrow1999}. 
We treat the data obtained during the caustic passage(s) separately 
from the remainder of the light curve using the criteria given below.

For a caustic crossing binary-lens event a minimum of $7+2n$  parameters are required, namely            
$t_0,u_0,t_{\rm{E}},q,d,\alpha$, and $\rho_\ast$, plus  $F_{\rm{S}}$ and $F_{\rm{B}}$ for each
of the $n$ different observing sites (here $n=6$). Here $t_0$ denotes the time of closest
approach to the center of mass of the binary, $u_0 \geq 0$ the impact parameter at time $t_0$, $t_E$ is the
time needed to cross the angular Einstein radius, which is defined as
\begin{equation}\label{eq:te}
\theta_{\rm{E}}=\sqrt{\frac{4GM}{c^2} \frac {D_{\rm{LS}}}{D_{\rm{L}} D_{\rm{S}}  }     }
\;,\end{equation} 
where $D_{\rm{L}},D_{\rm{S}}$ and $D_{\rm{LS}}$ are the observer-lens, observer-source, 
lens-source distances and $M$ the total mass of the binary lens.
The lens is characterized by the mass ratio $q=m_2/m_1$ between the
secondary and the primary and their angular separation
$d\,\theta_\mathrm{E}$. The impact angle $\alpha$ is measured between the
line from the secondary to the primary and the positive direction of
source motion relative to the lens. The angular source size is given
by $\rho_\star\,\theta_\mathrm{E}$.
The flux of the unlensed source star is $F_{\rm{S}}$ and $F_{\rm{B}}$ is the flux contribution 
of any other unlensed sources (including the lens) within the aperture. For every observing 
site, $F_{\rm{B}}$  and $F_{\rm{S}}$
are determined independently to account for different background and flux characteristics of the
individual telescopes/detectors. Modeling the parallax effect due to the orbital 
motion of the Earth requires 2 more parameters, the length $\pi_{\rm{E}}$ of the 
semi-major axis projected onto the sky plane and a rotation angle $\psi$, describing the relative 
orientation of the transverse motion of the source track to the ecliptic plane.
The source surface brightness profile in this study is described by either a 1- or 2-parameter law
so that the complete photometric model consists of up to 23 parameters.\\
Our initial search for the lens model involves only data outside the caustic-crossing 
region, where extended-source effects are negligible. Moreover, we also neglect
parallax effects.
We then scan the parameter space on a grid of fixed values of mass ratio $q$ and lens separation $d$, 
optimizing the remaining parameters $t_0,u_0,t_{\rm{E}}$ and $ \alpha$ with the
genetic algorithm Pikaia \citep{1995ApJS..101..309C,Kubas2004} and subsequently with a 
gradient routine to obtain $\chi^2$-maps such as shown in Fig.~\ref{fig:chi_scape_full}, 
which give an overview of possible model solutions. 
The values of $F_{\rm{S}}$ and $F_{\rm{B}}$ are simultaneously 
computed by  inexpensive linear fitting.
To explore in more detail the minima that are found we conduct a search with Pikaia over a restricted range 
of $q$ and $d$ but this time allowing these parameters to be optimized as well and again use gradient based
techniques for final refinement.
The results from the fold-caustic-crossing modeling in combination with the
point source fits are then used to generate magnification maps with the ray-shooting technique 
\citep{1997MNRAS.284..172W}. These maps contain the full information on the lens-source system.\\
We note that in crowded fields the raw photometry errors given by the reduction process clearly underestimate the
true errors \citep{2000AcA....50..421W}. To achieve a reduced $\chi^2$ of unity in our best fit model the photometric error 
bars would have to be rescaled by factors of 1.51 (SAAO), 1.92 (UTas), 1.34 (Danish), 1.16 (ESO 2.2m), 1.59 (Perth) 
and 2.3 (OGLE).
\subsection{Preferred lens parameters}\label{sec:psbl}
\begin{figure}
\includegraphics[width=9.cm]{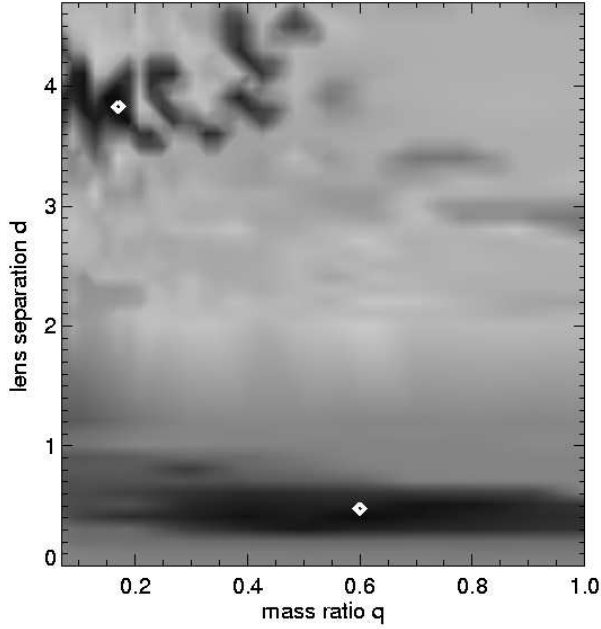}
\caption{Gray-scale $\chi^2$ map of full grid in mass ratio q and lens separation d,  
with darker regions representing lower values of $\chi^2$.
The irregularity of parameter space is reflected by the patchy appearance of the map, 
especially around the wide binary model in the upper left part of the plot. 
The best models are marked by the white diamonds at  
$ q\sim$ 0.6,~ $d \sim$ 0.5 (close binary)  and 
at $ q\sim$ 0.16, $d \sim$ 3.7 (wide binary).}
\label{fig:chi_scape_full}
\end{figure}

To exclude the data points which are affected by finite source effects we
apply the argument given in \cite{Albrow1999}. There it was shown, that for
times $\gtrsim 3\Delta t$ away from the fold caustic, where $\Delta t$ is 
the time in which the source radius crosses the caustic, the point source approximation 
is accurate enough for photometric errors of  $\lesssim 1\%$. Based on  the  measured 
caustic crossing times (see section \ref{sec:cc}) we cut out data between $2450.0\leq$ HJD'
$\leq 2452.8$ and $2463.0 \leq$ \mbox{HJD'} $\leq2466.0$, where $\mbox{HJD'}= \mbox{HJD}-2450000$. 
We then search for promising regions 
in parameter space on a grid 
of mass ratio $q$ and lens separation $d$, the two parameters that characterize the binary lens, with 
$q= 0.01, 0.05, 0.10, 0.15,..., 1.00$ and $d= 0.01, 0.05, 0.10, 0.15,..., 4.70$.
The result is shown in Fig.~\ref{fig:chi_scape_full}.
\psfull
While the apparent close binary solution around $ q\sim$ 0.6 and  
$d \sim$ 0.5 seems to be well defined, the numerical routines converge poorly in the vicinity of 
the wide-binary solution, reflecting the intricacy of binary-lens parameter space. By bracketing
apparently interesting subsets of the $(q,d)$ plane, our optimization algorithm identifies the 
best wide solution at $ q\sim$ 0.16 and $d \sim$ 3.7.
%
\begin{figure}
\includegraphics[width=8.8cm]{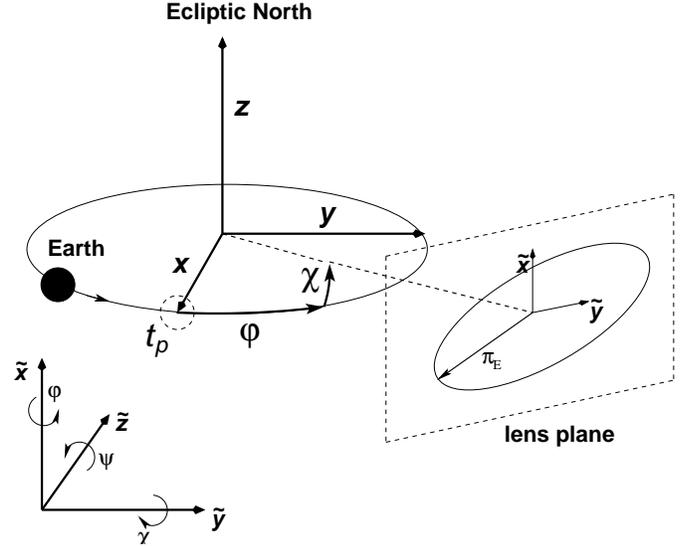}
\caption{The coordinates $(\tilde{x},\tilde{y})$
are chosen so that the right hand system $(\tilde{x},\tilde{y},\tilde{z})$ 
fulfils $\tilde{x}=z,\;\tilde{y}=y,\;\tilde{z}=-x$ for $\varphi=\chi=\psi=0$.
The longitude $\varphi$ is measured from the perihelion $t_{p}$ towards the Earth's motion and the
latitude $\chi$ from the ecliptic plane towards ecliptic north.}
\label{fig:parallax_convention}
\end{figure}

\subsection{Annual parallax}

Close-binary-lens models that neglect the motion of the Earth around the Sun show a significant asymmetry 
in the residuals which disappears if parallax is taken into account.
Adapting the convention in \cite{1998A&A...329..361D} and illustrated in Fig.~\ref{fig:parallax_convention} 
we introduce as a parameter 
the projected length $\pi_{\rm{E}}$ of the Earth's semi-major axis in the sky plane, which is defined as
\begin{equation}
\pi_\mathrm{E}
= \pi_{\mathrm{LS}}/\theta_{\mathrm{E}} =
\frac{1~{{\rm{AU}}}}{\theta_{\mathrm{E}}}\,
\left(\frac{1}{D_{\mathrm{L}}}-\frac{1}{D_{\mathrm{S}}}\right)\,,
\end{equation}
where $\pi_\mathrm{LS}$ is the relative lens-source parallax.
The second additional parameter is the angle $\psi$ describing the relative orientation 
of the source motion to the ecliptic. The heliocentric ecliptic coordinates $(\varphi,\chi)$ used for the
parallax modeling are derived from the standard geocentric ecliptic coordinates $(\lambda,\beta)$ 
by applying $\chi=\beta$ and $\varphi=\lambda+\pi+\varphi_\gamma$, where $\varphi_\gamma$ is the 
angle of the vernal equinox measured from the perihelion.\\
In 2002 Earth reached the perihelion at $\sim$ 2277.1 HJD' and the time of the vernal equinox
was $\sim$ 2354.3 HJD'. This yields $\varphi=163.3^\circ$ and $\chi=2.1^\circ$ for OGLE-BLG-2002-069.
%
\section{Source model}\label{sec:cc}

\begin{figure}
\includegraphics[width=9.cm]{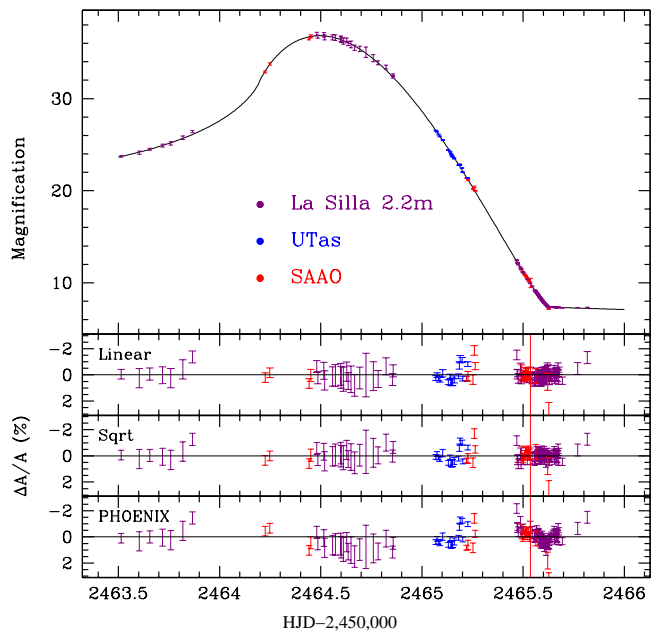}
\caption
    {
      Data points and light curve obtained with square-root limb darkening (upper panel), 
      and residuals coming from the linear, square-root and PHOENIX laws.
    }
\label{fig:LdFit}
\end{figure}

%
The data taken when the source transits the caustic show the corresponding characteristic shape.
The caustic entry is not well
sampled, because in the early stage of the event it was difficult to distinguish between
a binary and a single lens, and the practically unpredictable rise of the light curve was rather short. 
On the other hand, the caustic exit has very good coverage thanks to our predictive online modeling.
Hence, we focus our study on the caustic exit.
We estimate that the exit occurred for
($2463.45 \leq \mbox{HJD'} \leq 2467$) approximately. The corresponding subset of data 
comprises 95 points from ESO 2.2m, 21 points from SAAO and 17 points from
UTas, giving a total of 133 points.
We assume the source to move uniformly and neglect the 
curvature of the caustic as well as the variation of its strength on the 
scale of the source size. This approximation (which is justified in 
Sec. \ref{sec:complete}) allows us to increase computational efficiency significantly by using a
fold-caustic-crossing model \citep[e.g.][]{2004A&A...419L...1C}. 
We recall that during a caustic crossing, the total magnification $A^{(s)}$ of the source 
is the sum of the magnifications of the two critical images and the three other images :
\begin{equation}
  A^{(s)}= a_\mathrm{crit}\
  G_\mathrm{f}\left(\frac{t_\mathrm{f}-t}{\Delta t}\ ;\ \xi^{(s)}\right)
  + a_\mathrm{other}\ \left[1+\omega (t- t_\mathrm{f})\right].
\end{equation}
Here $\Delta t$ is the time needed for the radius of the source to cross
the caustic, $t_\mathrm{f}$ is the date at which the limb exits the caustic 
and $G_\mathrm{f}$ is a characteristic function
\citep{1987ApJ...314..154S} depending on the surface brightness profile $\xi$.
The blending parameters and the baseline magnitudes for each site are derived from the 
point-source model on the non-caustic-crossing part of the light curve ; they are held fixed in the
following.\\
Limb-darkening is frequently characterized by a sum of power-laws:
\begin{equation}
  \frac{\xi(\mu)}{\xi_0}=1-\sum\limits_i a_i (1-\mu^i),
\end{equation}
where $\mu=\cos\,\vartheta$ is the cosine of the emergent angle of the light ray
from the star, and $a_i~(i={1/2,1,3/2,2...}$) are the so-called
limb darkening coefficients (LDC).
We investigate the two most popular realizations : the linear ($a_1 \neq 0$) and square root 
limb darkenings ($a_1 \neq 0$ and $a_{1/2} \neq 0$).
Performing a $\chi^2$ minimization on our fold-caustic data 
provides us with the parameters listed in Table~\ref{tab:LDC} that best
describe our photometric data. 
The value of $\chi^2/\mathrm{d.o.f.}$ tells us about the relative goodness
of the fits among the studied models.
\cite{Claret2000} also introduced a 4-parameter law which fits the limb-darkening curves 
derived from spherical atmosphere models. 
However, as pointed out by \cite{2004MNRAS.352.1315D}, for coefficients beyond the linear law,
the differences in the light curve are much smaller than the differences in the profiles, and it is not
possible to find a unique set of coefficients given our data set.
Finally, we also consider a PHOENIX atmosphere model that resulted from a spectroscopic 
analysis of the source star by \cite{2004A&A...419L...1C}, where corresponding broad-band brightness 
profiles for $\mathrm{R}$- and $\mathrm{I}$-band were computed. \\
In the upper panel of Fig.~\ref{fig:LdFit}, the best model (with square-root limb darkening) is 
plotted with the data. The fit residuals obtained with the linear, square-root and PHOENIX
limb darkening are displayed in the lower panels.
With free limb-darkening coefficients,
even the linear law describes the data reasonably well, while the
square-root law allows a better match. In contrast the parameter-free
PHOENIX model computed assuming LTE fails. The residuals
for the caustic-crossing region show systematic trends that are typical
for an inappropriate limb-darkening profile, as discussed by
\cite{2004MNRAS.353..118D}.
A new analysis taking into account
NLTE effects will be done in a forthcoming paper.\\ 
\begin{table}
  \begin{center}
    \caption{Limb-darkening coefficients (LDC) derived from the fit of
OGLE-2002-BLG-069 during the caustic exit.
      Each set of LDC corresponds to the best fit obtained (no error
rescaling here) by using the LDC 
      as well as the fold-caustic model parameters as free parameters.
    }
    \begin{tabular}{|c|c|c|c|}
      \hline
       & Linear &  Square root & PHOENIX \\
      \hline
      $t_\mathrm{f}$ (days) & $2465.624$ & $2465.626$ & $2465.636$ \\
      \hline
      $\Delta t$ (days) & $0.71$ & $0.72$ & $0.72$ \\
      \hline
      ${\omega}$ (days$^{-1}$)  & $-0.10$ & $-0.10$ & $-0.11$ \\
      \hline
      $a_\mathrm{crit}$ & $19.98$ & $20.03$ & $19.85$ \\
      \hline
      $a_\mathrm{other}$ & $7.33$ & $7.35$ & $7.33$ \\
      \hline
      $a_1$ & $0.62$ & $0.10$ & - \\
      \hline
      $a_{1/2}$ & - & $0.80$ & - \\
      \hline
      $\chi^2/\mathrm{d.o.f.}$ & $2.230$ & $1.937$ &  $3.528$ \\
      \hline
    \end{tabular}
  \end{center}
  \label{tab:LDC}
\end{table}
In the following sections,
we will use the square-root limb darkening to describe the source star.

\section{A complete model}\label{sec:complete}
\begin{figure*}
\includegraphics[width=18.5cm]{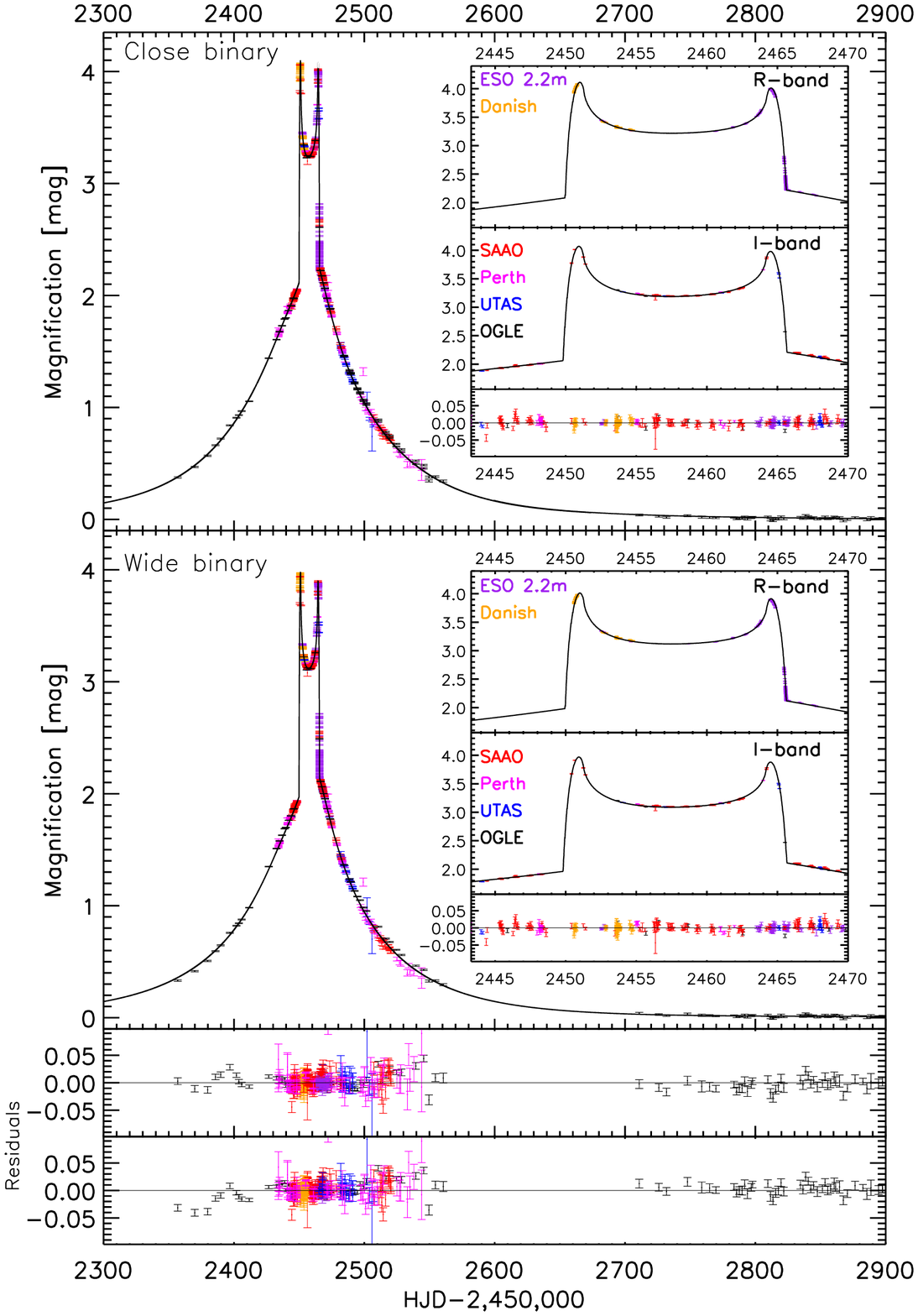}
\caption{Lightcurves of close and wide binary-lens models and their residuals. 
Insets show a zoom of the caustic passages in the two different bands taken.}
\label{fig:wide_light}
\end{figure*}

\begin{figure}
\includegraphics[width=9.cm]{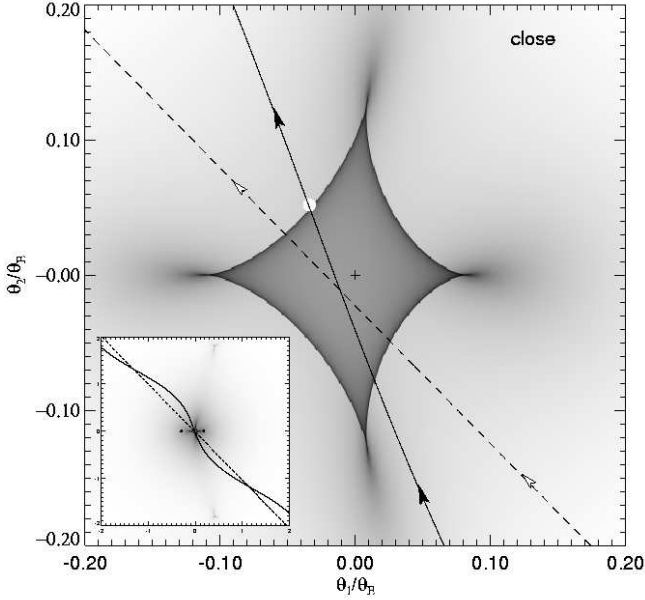}
\caption{Magnification map and source trajectory for the
close-binary model.
The solid curve is the track of 
the source including annual parallactic motion and the dashed line represents 
the source motion (with direction indicated by the arrows) as seen from the Sun. 
The origin marked with the cross is the center of mass and the filled 
white circle indicates the source size. The grey-scale marks the magnification scale 
in the source plane, with dark regions corresponding to high magnification and bright 
regions referring to low magnification. The inset shows the full caustic topology, with
the two filled black circles marking the positions of the binary lens components. }
\label{fig:close_maptrack}
\end{figure}

\begin{figure}
\includegraphics[width=9.cm]{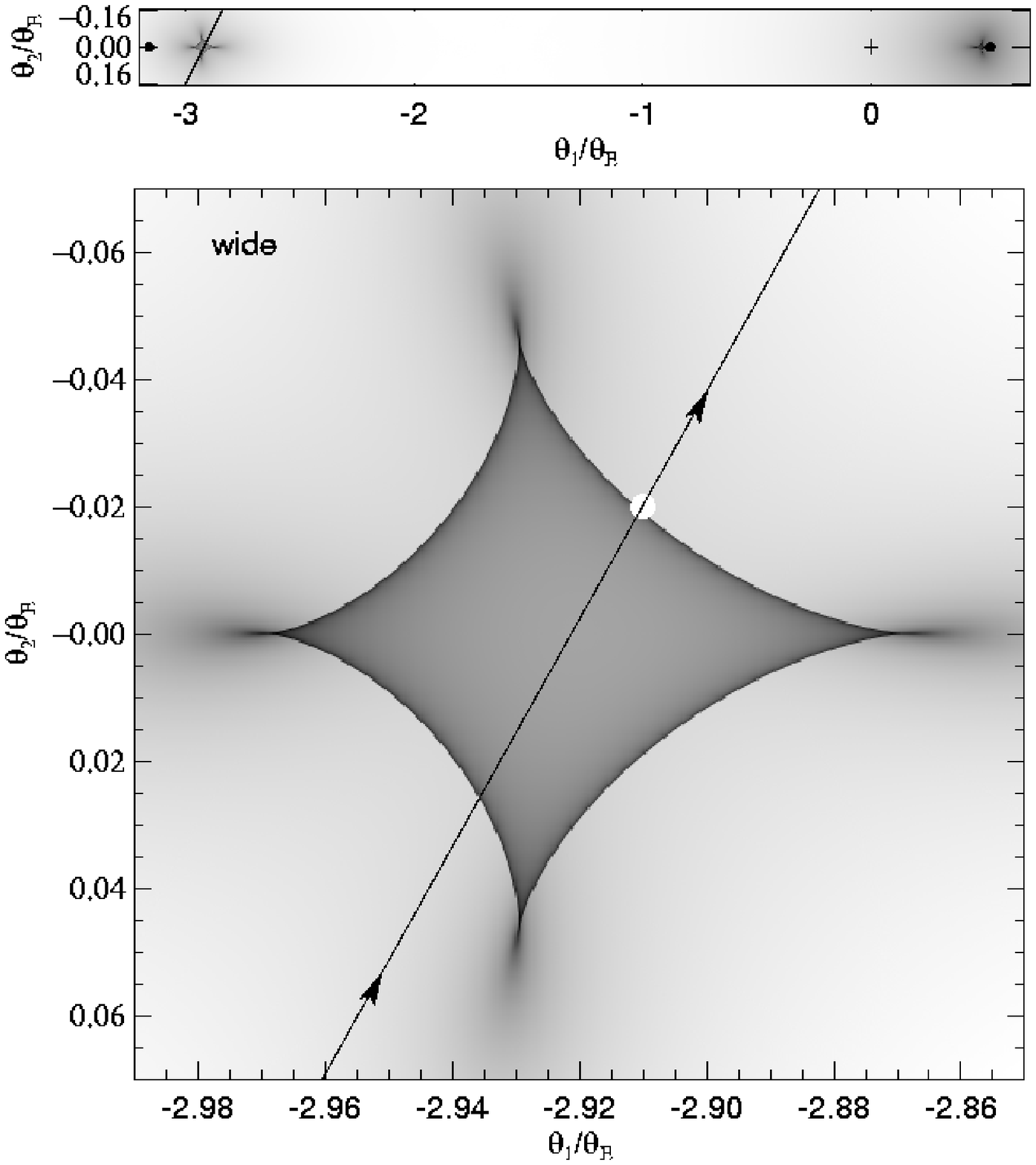}
\caption{ Magnification map and source trajectory for the
wide-binary. The effect of parallax is negligible. 
The top panel shows the full caustic topology, where the lens positions are
marked with the black filled circles and the center of mass by the cross.
In the zoom around the secondary lens,  {\bf{lower panel}}, the filled white circle marks the source size,
the arrows indicate the direction of source motion. As in Fig.~\ref{fig:close_maptrack} dark
regions mark high magnification and bright regions represent low magnification areas.}
\label{fig:wide_maptrack}
\end{figure}

With the point-source model and the brightness profile
of the source determined from the data in the caustic-crossing region, we
can now derive a complete and consistent model of the lens, yielding its
mass $M$, distance $D_{\rm{L}}$ and relative transverse velocity $v$.
This is done by generating magnification maps with the ray-shooting method \citep{1997MNRAS.284..172W} for
the best-fit values found for mass ratio $q$ and lens separation $d$ and then convolving these maps
with the source profile modeled in section~\ref{sec:cc}. 
The maps and the corresponding light curves derived from them,
are shown in figures~\ref{fig:wide_light},~\ref{fig:close_maptrack}, and~\ref{fig:wide_maptrack}.\\
These maps also serve as a  check on the validity of the straight-fold-caustic approximation 
(see Sec.~\ref{sec:cc}). We find that the effect of 
curvature of the caustic is negligible and does not influence the results of the
stellar surface brightness modeling.
Table \ref{parameter_tab} lists all fit parameters for the best close- and wide-binary solution. 
The quoted 1-$\sigma$ error bars correspond to projections of
the hypersurfaces defined by
$\Delta \chi^2 = \chi^2 - \chi^2_\mathrm{min} = 1$ onto the parameter axes.
%
%

%
%
%
\subsection{Physical lens properties}\label{sec:mass_det}
\begin{figure}
\includegraphics[width=9.0cm]{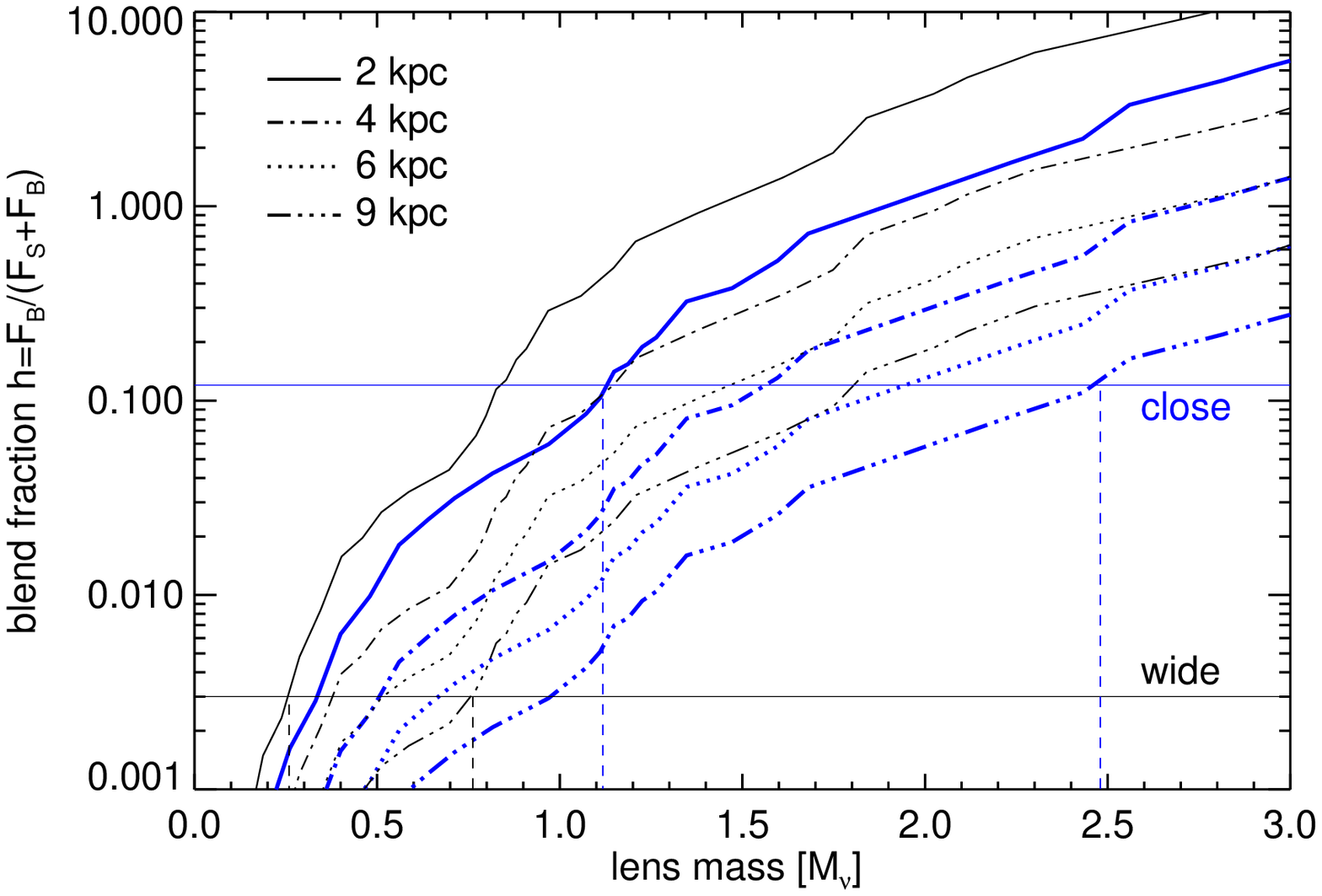}
\caption{Assuming the lens to be composed of two main-sequence stars according
to each of our models, the blend fraction 
$h= F_{\rm{B}}/(F_{\rm{S}}+F_{\rm{B}})$ for OGLE data (horizontal solid lines) 
yields upper limits (vertical dashed lines) on the total lens mass for the known 
brightness and distance of the source ($M_V=+0.9$, $(V-I)=0.95)$ and 
$D_{\rm{S}}=(9.4\pm1.4)$ kpc from \cite{2004A&A...419L...1C}. While the close-binary model 
is compatible with lens masses
up to  $\sim 1.1-2.5\; {\rm{M}}_\odot$, the wide-binary model only allows lens masses up to
$\sim 0.3-0.8\; {\rm{M}}_\odot$ for lens distances between 2-9 kpc (where the thin lines represent the
wide binary lens and the thick lines the close binary lens). }
\label{fig:luminosity_contraints}
\end{figure}

The measured finite source size and the parallax effect yield two  independent constraints for 
determining the lens mass $M$, its distance $D_{\rm{L}}$ and transverse velocity $v$. 
Assuming a luminous lens we can put upper limits on its mass using our knowledge of the absolute 
luminosity and distance of the source star. These were determined in \cite{2004A&A...419L...1C}
from spectroscopic measurements combined with the measured  amount of blended light (which includes 
any light from the lens) inferred from the light curve modeling.\\
Fig.~\ref{fig:luminosity_contraints} plots the implied
blend fraction $h= F_{\rm{B}}/(F_{\rm{S}}+F_{\rm{B}})$ if both components of the lens are  
main-sequence stars \cite[from A0 to M9,][]{allen} put at distances of $2,4,6$, and $9$ kpc along the
line of sight to the lens in comparison with the blend fractions derived from OGLE data. If we assume
the lens is the only source of the blended light,
the inferred blend fraction from our best fit models gives an upper limit for the total lens mass of 
$\sim 2.5 ~{\rm{M}}_\odot$ for the close-binary and $\sim 0.8 ~{\rm{M}}_\odot$ for the wide-binary-lens model.\\ 
%
\begin{figure}
\includegraphics[width=9.0cm]{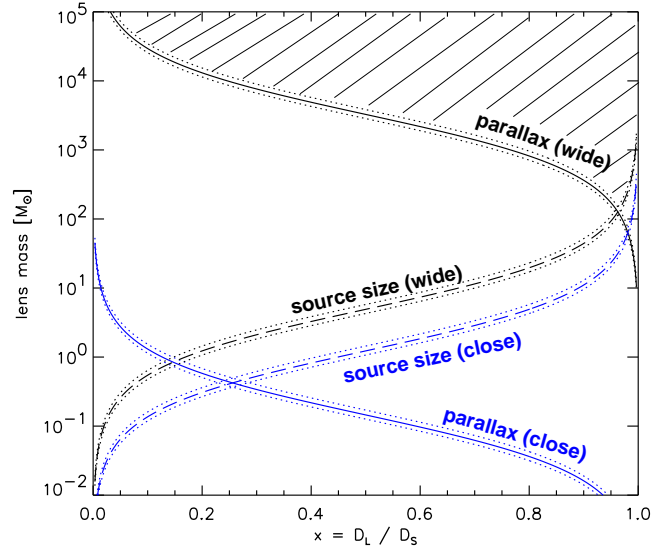}
\caption{The independent constraints on the lens mass from the source size (dashed line) 
according to Eq.~(\ref{eq:mass_x}) and from parallax effects according to Eq.~(\ref{eq:mass_eta}) (solid line) 
for the close and wide separation binary models as function
of $x=D_{\rm{L}}/D_{\rm{S}}$. The dotted lines mark the uncertainty due to the error 
in the source distance measurement. While the close-binary constraints intersect 
at the plausible lens mass of $(0.51 \pm 0.15) ~{\rm{M}}_\odot$ implying 
a disc lens at $D_{\rm{L}}=(2.9\pm 0.4) ~{\rm{kpc}}$, the wide binary favours a 
rather implausible scenario of a binary black hole of $M\gtrsim (126 \pm 22) ~M_\odot$ in the 
Bulge at $D_{\rm{L}}\gtrsim (9.0 \pm 2.3) ~{\rm{kpc}}$, with the hatched region marking the
allowed parameter space for the wide-binary solution.}
\label{fig:xm_contraints}
\end{figure}    
%
%
We use the source radius caustic crossing time $\Delta t$ from the straight-fold-caustic model, 
together with the lens geometry given by $(q,d)$  and time-scale $t_{\rm{E}}$
from the point source model, to derive the relative angular source size $\rho_\ast$, which
expressed as a fraction of the angular Einstein radius reads  
\begin{equation}\label{eq:rho}
\rho_{\ast}=\frac{\Delta t} {t_{\rm{E}}}\;\rm{sin} \phi
\;,
\end{equation}
with $\phi$ being the angle between source track and caustic tangent. 
The source size parameter $\rho_\ast$ is refined by fitting on a grid
of magnification maps convolved with different source sizes.
So with the inferred physical source size of  $ R_{\ast}\simeq 10~R_{\odot}$  and source 
distance $D_{\rm{S}}=(9.4\pm 1.4)\;\rm{kpc}$ from the spectral measurements
\citep{2004A&A...419L...1C} the constraint on the lens mass from extended-source effects 
can be inferred from 
\begin{equation}\label{eq:mass_x}
\frac{M(x)}{M_{\odot}}=\frac{{c^2}} {4GM_{\odot} D_{\rm{S}}} 
\frac {{R_{\ast}}^2}{{\rho_\ast}^2}  \frac{x}{1-x}
\;,\end{equation}
with $x=D_{\rm{L}}/D_{\rm{S}}$.
The dependence of the lens mass upon annual parallax effects reads 
\begin{equation}\label{eq:mass_eta}
\frac{M(x)}{M_{\odot}}=\frac{c^2}{4GM_{\odot}D_{\rm{S}}} \left(\frac{1~{\rm{AU}}}{\pi_{\rm{E}}}\right)^2 \frac{1-x}{x}~~.
\end{equation}
The curves arising from these two relations are plotted in Fig.~\ref{fig:luminosity_contraints} for our best fit 
parameters $\pi_{\rm{E}},\rho_\ast$ (see Table \ref{parameter_tab}) of the wide and close binary-lens model. From this we obtain
the following physical lens parameters,
\begin{equation}\label{eq:lensmass_close}
M_{\rm{close}}=(0.51 \pm 0.15) ~M_\odot\;,
\;{\rm{at}}\;  
D_{\rm{L}}=(2.9 \pm 0.4)~\rm{kpc}\;
\end{equation}
and 
\begin{equation}\label{eq:lensmass_wide}
M_{\rm{wide}}\gtrsim (126 \pm 22) ~M_\odot\;,
\;{\rm{at}}\;  
D_{\rm{L}}\gtrsim (9.0 \pm 2.3) ~\rm{kpc}\;.
\end{equation}
The close-binary solution yields a Bulge-Disc lens scenario similar to that for
EROS-BLG-2000-5 \citep{2002ApJ...572..521A}, namely an M-dwarf binary system with a projected
separation of $(4.5 \pm 1.1) \;\rm{AU}$ located most likely just beyond the Orion arm of the Milky Way. 
The marginal detection of parallax effects in the wide-binary model however allows us to put lower limits 
on the mass and velocity of the lens, suggesting a rather implausible binary system consisting of two 
super-stellar massive black holes in the Galactic bulge with $v_{\rm{wide}}\gtrsim 129~\rm{km}~\rm{s}^{-1}$. 
We therefore reject the wide binary model and derive for the transverse velocity  of the close 
binary model
\begin{equation}\label{eq:lensveloc}
v_{\rm{close}}=\frac{R_\ast} {t_{\rm{E}}(\frac{\pi_{\rm{E}}}{{1~\rm{AU}}}\; R_\ast+\rho_\ast)}=(49.8 \pm 2.7)~ \rm{km}~\rm{s}^{-1}\;.
\end{equation}

\begin{table}[h!]
\begin{center}
\begin{tabular}{|l|c|c|}  \hline
parameter & close & wide  \\\hline
& &   \\
$ q$& $0.58^{+0.05}_{-0.02}$ & $0.17^{+0.06}_{-0.02}$       \\   
$d$ & $0.46^{+0.01}_{-0.02}$  &  $3.68^{+0.02}_{-0.02}$    \\
$u_0$ & $0.016^{+0.001}_{-0.003} $    & $2.562^{+0.005}_{-0.002}$     \\
$\alpha$ [deg]& $134.4^{+0.5}_{-0.5}$     &   $60.8^{+0.5}_{-0.5}$\\
$t_{\rm{E}}$ [days]& $104.5^{+1.0}_{-1.0}$        & $286.5^{+3.0}_{-2.0}$  \\
$t_{0}$ & $2456.92^{+0.10}_{-0.10}$        & $2825.32^{+0.12}_{-0.12}$  \\
$\pi_{\rm{E}}$    &$ 0.23^{+0.02}_{-0.02}$     & $0.002^{+0.05}_{-0.02}$\\
$\psi$ [deg]   &$ 258.551^{+12.2}_{-12.2}$    & $189.0^{+10.5}_{-10.5}$ \\
$\rho_\ast$  & $0.0048^{+0.0005}_{-0.0005}$      & $0.0023 ^{+0.0009}_{-0.0009}$\\
$h_{\rm{\tiny{OGLE}}}$   & $0.10^{+0.02}_{-0.02}$      & $0.00^{+0.03}_{-0.00}$\\\hline
$\chi^2/{\rm d.o.f.}$& 2029.8 / 631    & 2251.0 / 631 \\\hline
\end{tabular}
\caption{Fit parameters of best close and wide-binary models with $1\sigma$ uncertainties.
The $\chi^2$ values are based on the raw photometric errors.
i.e. without rescaling factors.}
\label{parameter_tab}
\end{center}
\end{table}
%
%
\section{Summary and Conclusions} \label{sec:disc}
%
While the number of observed galactic microlensing events has now reached an impressive
count of over 2000 (with about 5 \% of them being identified as binary-lens events), 
still very little is known about the physical properties of the 
lens population, since in general the information on mass, distance and velocity of
the lens needs to be inferred from one single parameter, the event time scale $t_{\rm{E}}$. 
The present work is the second successful attempt \citep[after][] {An2002} at putting strong
constraints on lens and source properties in a microlensing event. This event involves a G5III cool giant
in the Bulge at a distance of $D_{\rm{S}}=(9.4\pm 1.4)\;\rm{kpc}$ lensed by an M-dwarf 
binary system of total mass $M_{\rm{}}=(0.51 \pm 0.15) ~M_\odot$ located at 
$D_{\rm{L}}=(2.9 \pm 0.4) \;\rm{kpc}$. These conclusions could only be achieved by the use of a network
of telescopes to ensure a continues, dense and precise coverage of the event, whereas data obtained
from a survey with mainly daily sampling are insufficient for achieving this goal \citep{2004AcA....54..103J}.
The parameter space exploration, for both lens and source properties, described here provides 
a template for our future analysis of binary-lens events with fold-caustic crossings.
%
\begin{acknowledgements}
%
The Planet team wishes to thank the OGLE collaboration for its
fast Early Warning System (EWS) which provides a large fraction of the targets for
our follow-up observations. 
Futhermore we are especially grateful to the observatories that support our
science (European Southern Observatory, Canopus, CTIO, Perth, SAAO)
via the generous allocation of telescope time that makes this work possible. 
The operation of Canopus Observatory is in part supported by the financial
contribution from David Warren, and the Danish telescope at La Silla
is operated by IDA financed by SNF.  JPB acknowledges
financial support via an award from the ``Action Th\'{e}matique Innovante''
INSU/CNRS. MD acknowledges postdoctoral support on the PPARC rolling grant 
PPA/G/2001/00475.
\end{acknowledgements}
\bibliographystyle{aa}


\end{document}